\documentclass[prd,aps,superscriptaddress,nofootinbib,twocolumn,a4paper,preprintnumbers]{revtex4}
\usepackage{amsmath,amssymb,amsfonts,color}
\usepackage{bm}
\usepackage{latexsym}
\usepackage[utf8]{inputenc}

\usepackage{hyperref}
\DeclareMathAlphabet{\mathpzc}{OT1}{pzc}{m}{it}

\definecolor{brown}{rgb}{0.65,0.16,0.16}

\usepackage{xspace}
\newcommand{\aether}{\ae{}ther\xspace} 

\begin{document}
\preprint{Imperial/TP/2016/AEG/5}
\title{Revisiting the cuscuton as a Lorentz-violating gravity theory}
\author{Jishnu Bhattacharyya}
\affiliation{School of Mathematical Sciences, University of Nottingham, University Park, Nottingham, NG7 2RD, United Kingdom}
\author{Andrew Coates}
\affiliation{School of Mathematical Sciences, University of Nottingham, University Park, Nottingham, NG7 2RD, United Kingdom}
\author{Mattia Colombo}
\affiliation{School of Mathematical Sciences, University of Nottingham, University Park, Nottingham, NG7 2RD, United Kingdom}
\author{A.~Emir G\"umr\"uk\c{c}\"uo\u{g}lu}
\affiliation{Theoretical Physics Group, Blackett Laboratory, Imperial College London, South Kensington Campus, London, SW7 2AZ, UK}
\author{Thomas P. Sotiriou}
\affiliation{School of Mathematical Sciences, University of Nottingham, University Park, Nottingham, NG7 2RD, United Kingdom}
\affiliation{School of Physics and Astronomy, University of Nottingham, University Park, Nottingham, NG7 2RD, United Kingdom}

\date{\today}
\begin{abstract}
The cuscuton is a scalar field with infinite speed of propagation. It was introduced in the context of cosmology but it has also been claimed to resemble Ho\v rava gravity in a certain limit. Here we revisit the cuscuton theory as a Lorentz-violating gravity theory. We clarify its relation with Ho\v rava gravity and  Einstein-\aether theory, analyze its causal structure, and consider its initial value formulation. Finally, we discuss to which extent the cuscuton theory can be used as a proxy for Ho\v rava gravity in the context of gravitational collapse and formation of universal horizons.
\end{abstract}
\maketitle
\section{Introduction}

Lorentz symmetry appears to be a fundamental symmetry of the Standard Model of particle physics and experimental constraints are impressively tight (see {\em e.g.}~\cite{Mattingly:2005re}). Conversely, gravitational experiments are significantly less sensitive to Lorentz violations. A way to make this statement quantitative is to consider Lorentz-violating alternatives to general relativity and confront their predictions with observations. Einstein-\aether theory \cite{Jacobson:2000xp}  is perhaps the most well-studied theory in this category. Its action is 
\begin{equation}
\label{aeaction}
S_{\text{\ae}}=\frac{M_{pl}^2}{2}\,\int \mathrm{d}^4 x \sqrt{-g}\left({}^{(4)}R-M^{\mu\nu}{}_{\alpha\beta}\nabla_\mu u^\alpha \nabla_\nu u^\beta\right)\,,
\end{equation}
where \({}^{(4)}R\) is the 4-dimensional Ricci scalar and
\begin{equation}
M^{\mu\nu}{}_{\alpha\beta} = c_1 g^{\mu\nu}g_{\alpha\beta}+c_2 \delta^{\mu}_{\alpha}\delta^{\nu}_{\beta}+c_3 \delta^{\mu}_{\beta}\delta^{\nu}_{\alpha}-c_4 u^\mu u^\nu g_{\alpha\beta}\,.
\end{equation}
The vector $u^\mu$, dubbed the \aether, is constrained to be unit timelike, {\em i.e.}~$u^\mu u^\nu g_{\mu\nu}=-1$ when using the metric signature \((-,+,+,+)\),  either by using a Lagrange multiplier or by constraining the \aether's variation. Due to this constraint, the \aether can never vanish and it always defines preferred timelike trajectories in spacetime (a preferred threading''). This violates local Lorentz invariance as, even in a local coordinate frame where the metric can be taken as flat in a neighborhood of some event, the \aether will define some preferred time direction. 

Another Lorentz-violating theory that has attracted a lot of attention recently is Ho\v rava gravity \cite{Horava:2009uw}. This is a theory with a preferred foliation. In a covariant form the low energy limit of Ho\v rava gravity can be considered dynamically equivalent to Einstein-\aether theory with the additional restriction 
 \begin{equation}
 \label{aescalar}
u_\mu = -\frac{\partial_\mu T}{\sqrt{-g^{\mu\nu}\partial_\mu T \partial_\nu T}},
\end{equation}
where $T$ is a scalar field \cite{Jacobson:2010mx}. In this formulation the level surfaces of $T$ define the preferred foliation. Once this foliation has been adopted, the residual symmetries are diffeomorphisms that preserve it, {\em i.e.}~$t \to \tilde{t}(t)$ and $x^i \to \tilde{x}^i (t, x^i)$. In the most general version of the theory \cite{Blas:2009qj} the full action contains all additional operators compatible with this symmetry that have up to six spatial derivatives in the preferred foliation. Including these operators improves the behavior of the propagators in the ultraviolet and renders the theory power-counting renormalizable. The possibility of trading Lorentz-symmetry for improved ultraviolet behavior has certainly provided new-found motivation for considering Lorentz-violating theories of gravity. Here however, we will focus mostly on the low energy limit of the theory, so the covariant description given by the action \eqref{aeaction} together with the definition \eqref{aescalar} that identifies the field content will suffice.\footnote{For other restricted or extended versions of Ho\v{r}ava gravity, which will not be considered here, please see Refs.~\cite{Horava:2009uw,Sotiriou:2009gy,Sotiriou:2009bx,Weinfurtner:2010hz,Mukohyama:2010xz,Sotiriou:2010wn,Horava:2010zj,Sotiriou:2011dr,Vernieri:2011aa,Vernieri:2012ms,Colombo:2014lta,Colombo:2015yha,Coates:2016zvg}.}

The third theory we will discuss  is cuscuton theory \cite{Afshordi:2006ad}, which will be the main focus of this paper. The action of the theory is 
\begin{equation}
\label{cuscaction}
S=\int \mathrm{d}^4 x\sqrt{-g}\left(\frac{M_{pl}^2}{2}\,{}^{(4)}R+\mu^2\sqrt{2X}-V(\phi)\right).
\end{equation}
where, \(\phi\) is the cuscuton field,
 \begin{equation}\label{Xdefinition}
X=-\frac{1}{2} g^{\mu\nu}\partial_\mu\phi\partial_\nu\phi\,,
\end{equation}
 and \(V(\phi)\) is the cuscuton's potential. The presence of the square root in the kinetic term of the cuscuton straightforwardly implies that \(X\) must be nonnegative. Variation with respect to $\phi$ will actually lead to contributions in the cuscuton's field equation that contains $ 1/\sqrt{X}$. Hence, $X$ should be strictly positive and the cuscuton is forced to always have a timelike gradient. Therefore, the level surfaces of $\phi$ define a special foliation and the cuscuton is known to propagate with infinite speed \cite{Afshordi:2006ad}.
 
 The cuscuton was initially introduced in the context of cosmology \cite{Afshordi:2006ad} but the existence of a special foliation suggests some similarity with Ho\v rava gravity. In fact it has been claimed that the cuscuton acts as a low energy limit of Ho\v rava gravity \cite{Afshordi:2009tt}. Our main goal is to revisit this claim and to establish that this special foliation acts as a preferred foliation from a perspective of causality ({\em i.e.}~it determines the causal structure along the lines discussed in Ref.~\cite{Bhattacharyya:2015gwa}).\footnote{The term``preferred" frame or foliation is generically used in the Lorentz symmetry breaking literature in the sense of distinguishable or detectable. However, there is also the question of whether a frame or a foliation is preferred in the sense of being the unique sensible choice to, {\em e.g.},~discuss causality, set up an evolution problem or interpret results physically.}

 Our motivation is twofold. Firstly, a quick and na\"ive qualitative comparison between the cuscuton and the low energy limit of Ho\v rava gravity leads to  a puzzling observation. Action \eqref{aeaction} with $u_\mu$ given by Eq.~\eqref{aescalar} contains more than two derivatives on $T$ and variation with $T$ leads to a fourth order differential equation. Choosing $T$ as a time coordinate reduces the number of time derivatives to two \cite{Jacobson:2010mx} and this is precisely what singles out the foliation defined by $T=$ constant hypersurfaces as preferred. On the other hand, the cuscuton's equations are second order in derivatives in any foliation in the first place. Hence, if the cuscuton's foliation is preferred in some way, this should come about in a more subtle way. We will clarify this fully below and we will present a first discussion of the initial value problem in cuscuton theory. 
 
 The second piece of motivation comes from the fact that the cuscuton has been used as a proxy theory to understand gravitational collapse and the dynamical formation of {\em universal horizons} therein \cite{Saravani:2013kva}. In theories with infinite propagation, universal horizons \cite{Barausse:2011pu,Blas:2011ni} play the same role that event horizons have in general relativity. They act as a causal boundary for all excitations and hence they are the defining characteristic of a black hole. Whether and how they form during gravitational collapse is still an open problem in Ho\v rava gravity. To which extent the study of dynamical formation of universal horizon in cuscuton theory in Ref.~\cite{Saravani:2013kva} can teach us something about universal horizon formation in Ho\v rava gravity hinges strongly on the relation between the two theories and their causal structure.

\section{Reformulation of the cuscuton}

We start by reviewing a reformulation of the cuscuton as a scalar-vector-tensor theory following~\cite{Afshordi:2009tt}. This reformulation will prove particularly useful for our purposes. Consider the action
\begin{align}
\label{svtaction}
S[\phi,u,g] = \int\mathrm{d}^4x\sqrt{-g}\left[\frac{M_{pl}^2}{2}\,{}^{(4)}R+\mu^2\left(u^\mu\nabla_\mu\phi\right)\right.\nonumber \\
\left.\vphantom{{}^{(4)}R+\mu^2\left(u^\mu\nabla_\mu\phi\right)-V(\phi)}-V(\phi)+\frac{M_{pl}^2}{2}\sigma\left(g_{\mu\nu}u^\mu u^\nu+1\right)\right],
\end{align}
where \(\sigma\) is a Lagrange multiplier that enforces the unit-timelike-norm constraint for the vector field,~\textit{i.e.},
	\begin{equation}
	g_{\mu \nu} u^{\mu} u^{\nu} = -1~.
	\end{equation}
The equations of motion one gets through variation with respect to $g^{\mu\nu}$, $u^\mu$ and $\phi$ respectively are
\begin{align}
\label{equationsofmotion:1}
E_{\mu\nu} &\equiv G_{\mu\nu} - \frac{T_{\mu\nu}}{M_{pl}^2}=0~, \\
\label{equationsofmotion:2}
F_{\mu} &\equiv M_{pl}^2\sigma \,u_\mu +\mu^2\nabla_\mu\phi=0~,\\
\label{equationsofmotion:3}
H &\equiv -\mu^2 \theta -\frac{\mathrm{d}V(\phi)}{\mathrm{d}\phi}=0~,
\end{align}
where we have used the notation, \(\theta \equiv \nabla_\mu u^\mu\). In Eq.~\eqref{equationsofmotion:1} above, we defined the stress-energy tensor by $T_{\mu\nu} \equiv -(2/\sqrt{-g}) \delta \sqrt{-g}{\cal L}[\phi,u]/\delta g^{\mu\nu}$, where the Lagrangian density $\mathcal{L}[\phi, u]$ contains all term in action~\eqref{svtaction} apart from the Einstein-Hilbert term. The stress tensor reads
\begin{equation}\label{stressenergy}
T_{\mu\nu}=\left[\mu^2u^\alpha\nabla_\alpha\phi-V(\phi)\right]g_{\mu\nu}+M_{pl}^2\sigma \,u_\mu u_\nu.
\end{equation}
The vector field $u^\mu$ can be thought of as an auxiliary field,~\textit{i.e.}~its equation of motion can be solved algebraically for $u^\mu$.  Solving the vector's equation of motion and inserting the result into the action along with the unit constraint yields the cuscuton action \eqref{cuscaction}, provided that \(\sigma>0\). Note that the field equations above imply $X=M_{pl}^4\sigma^2/(2\,\mu^4)$ so positivity of $X$ is guaranteed. The \(\sigma <0\) branch corresponds to an analytic continuation of $\mu$ to the imaginary plane. In simpler words, choosing this branch is equivalent to flipping the sign of the $\mu^2\sqrt{2X}$ term in action \eqref{cuscaction}. It is worth emphasising that there does not seem to be a particular reason to restrict the sign of this term, or equivalently $\mu^2$, to be positive.  We have simply maintained the notation of Ref.~\cite{Afshordi:2006ad} for straightforward comparison.


\section{Formal equivalence to a special class of generalized Einstein-\aether theories}
\label{formaleq}

\subsection{General potential}
\label{sec:general-potential}
We have established that the scalar-vector-tensor theory above is equivalent to  cuscuton theory. However, a straightforward integration by parts on the second term in action \eqref{svtaction} would actually make $\phi$, instead of $u^\mu$, an auxiliary field. Indeed, {\em assuming that \(V(\phi)\) is not constant or linear in \(\phi\)} and that \textit{it can be formally inverted}, one has
	\begin{equation}\label{theta:phi:inversion}
	\theta = -\frac{1}{\mu^2}\frac{\mathrm{d}V(\phi)}{\mathrm{d}\phi} \qquad\leftrightarrow\qquad \phi = \frac{M_{pl}^2}{2\,\mu^2}\frac{\mathrm{d}\Phi(\theta)}{\mathrm{d}\theta}~.
	\end{equation}
The particular form of the inversion given above has been deliberately chosen such that the functions $V(\phi)$ and $\Phi(\theta)$ are related in the following way:
\begin{equation}\label{def:Phi:phi}
\mu^2 \phi \,\theta + V(\phi) = \frac{M_{pl}^2}{2}\Phi(\theta)~.
\end{equation}
Hence, once $\phi$ has been eliminated, the action~\eqref{svtaction} can be recast as
\begin{equation}
\label{genaetheory}
S[u,g]=\frac{M_{pl}^2}{2}\int\mathrm{d}^4x\sqrt{-g}\left[{}^{(4)}R-\Phi(\theta)+\sigma\left(g_{\mu\nu}u^\mu u^\nu+1\right)\right].
\end{equation}
This is exactly the form of the ``generalized Einstein-\aether theory" studied in Ref.~\cite{Zlosnik:2006zu}.  

The equation of motion for the vector that one obtains from this action is,
\begin{equation}
\label{genaether}
\frac{\mathrm{d}^2\Phi}{\mathrm{d}\theta^2}\nabla_\mu \theta+2\sigma u_\mu=0.
\end{equation}
From the form of the equation it follows that on-shell, $u_\mu$ is hypersurface orthogonal provided \(\sigma \neq 0\). This is expected, as cuscuton theory does not propagate any vector modes. Note that solutions with $\sigma=0$  exist if one takes action \eqref{genaetheory} at face value, but they will not be solutions of the cuscuton theory as the equivalence breaks down when $\sigma = 0$. Nonetheless we will consider these solutions in a bit more detail below. 

It is convenient to project Eq.~\eqref{genaether} along $u^\mu$ and normal to $u^\mu$. Contracting with the \aether and the projector $h^\mu_{\;\;\nu}= \delta^\mu_\nu+u^\mu u_\nu$, one gets
\begin{align}
\label{ugenae}
\frac{\mathrm{d}^2\Phi}{\mathrm{d}\theta^2}u^\mu\nabla_\mu \theta &=2\sigma,\\
\label{hgenae}
\frac{\mathrm{d}^2\Phi}{\mathrm{d}\theta^2} h^\mu_{\;\;\nu} \nabla_\mu \theta &=0.
\end{align}
As long as $\sigma\neq 0$, $u_\mu$ is hypersurface orthogonal by virtue of Eq.~\eqref{genaether} and hence the surface to which $u^\mu$ is normal defines a foliation. Then Eq.~\eqref{hgenae} implies that the leaves of this foliation have constant mean curvature $\theta$  and the value of $\theta$ on each leaf with depend on boundary conditions. Eq.~\eqref{ugenae} can then be used to determine $\sigma$ from the rate of change of $\theta$ as one moves through the leaves. 

If instead $\sigma=0$ then Eq.~\eqref{genaether} no longer forces $u_\mu$ to be hypersurface orthogonal. Hence there are two types of solutions.  There are solutions of the type where $u_\mu$ is not hypersurface orthogonal and Eq.~\eqref{genaether} implies that $u^\mu$ has constant covariant divergence. This is not enough to fully determine $u^\mu$, even when the constraint $u^\mu u^\nu g_{\mu\nu}=-1$ is taken into account. Moreover, it does not seem likely that these solutions are dynamically connected to the $\sigma\neq 0$ solutions. Hence, they are not particularly interesting. The second type of solutions are those where $u_\mu$ is ``accidentally'' hypersurface orthogonal. These are solutions were $u_\mu$ defines a foliation with leaves of constant mean curvature throughout.

The ``accidental'' hypersurface orthogonality might seem suspicious, but there is at least one case where it might be justifiable: stationary solutions. Let \(\chi^\mu\) be a Killing vector that is timelike in some region and assume that $u^\mu$ respects the corresponding symmetry. Then $\theta = \nabla_\mu u^\mu$ will also respect the Killing symmetry,~\textit{i.e.}
	\begin{equation}
	\mathcal{L}_\chi \theta = \chi^\mu \nabla_\mu \theta = 0~.
	\end{equation}
Eq.~\eqref{genaether} projected along $\chi^{\mu}$ then reads,
\begin{equation}
2\sigma \chi^\mu u_\mu =0.
\end{equation}
Since $u^\mu$ is timelike everywhere, in a region where $\chi^\mu$  is also timelike \(\sigma\) must be zero. Hence, stationary solutions are $\sigma = 0$ solutions. If such solutions arise as endpoints in the evolution of $\sigma\neq 0$ solutions, then continuity is enough to guarantee that $u^\mu$ will be hypersurface orthogonal.

The above discussion raises doubts about whether the cuscuton admits stationary solutions, as when \(\sigma =0\) the equivalence with the cuscuton fails. However, one can resort directly to the cuscuton's equation of motion
	\begin{equation}\label{scalarequation}
	\nabla_\mu\left(\frac{\nabla^\mu\phi}{\sqrt{2X}}\right) = \frac{1}{\mu^2}\frac{\mathrm{d}V(\phi)}{\mathrm{d}\phi},
	\end{equation}
in order to answer this question. The presence of $X$ in the denominator and under a square root implies that the cuscuton's gradient has to be timelike. Hence, $\chi^\mu\nabla_\mu \phi \neq 0$ in a region where $\chi^\mu$ is timelike. Therefore, $\phi$ cannot respect stationarity. Since $\phi$ appears without derivatives in the action and field equations due to the potential term, stationary solutions cannot exist in general. An exception might be the case where $V(\phi) = $ constant and one is willing to restrict stationarity to the metric only, but allow the cuscuton to have, say, linear dependence on Killing time.

\subsection{Quadratic potential}

The case of a quadratic potential, $V(\phi)=\alpha \phi^2$, is not qualitatively different from the general case. However, it is formally equivalent to a corner of Einstein-\ae ther theory,
albeit a highly-unrepresentative one. Direct substitution into Eq.~\eqref{theta:phi:inversion} yields $\phi=-\mu^2 \theta/(2\alpha)$, and hence action \eqref{genaetheory} can be rewritten as 
\begin{equation}
S[u,g]=\frac{M_{pl}^2}{2}\int\mathrm{d}^4x\sqrt{-g}\left[{}^{(4)}R-c_2 \theta^2+\sigma\left(g_{\mu\nu}u^\mu u^\nu+1\right)\right],
\end{equation}
where $c_2\equiv -\mu^4/(2\alpha\,M_{pl}^2)$. This is the action of Einstein-\aether theory with $c_1 = c_3 = c_4 = 0$.

The reason that this choice of parameters is special and nonrepresentative of generic Einstein-\aether theory should already be clear from the discussion after Eq.~\eqref{hgenae}: the vector field is either hypersurface orthogonal or underdetermined. In the former case, the \aether equation can actually be split into two parts, one of which requires boundary conditions in order to be solved. This is an indication that there is an instantaneous (elliptic) mode.\footnote{We call  elliptic a mode that satisfies an elliptic equation that is not a constraint, {\em i.e.}~it is not preserved by time evolution and it does not relate only initial data.}  Another way to see that this corner of the parameter space is special is to recall that the speeds of the various modes in Einstein-\aether theory are \cite{Jacobson:2004ts}
\begin{align}
c_{s(2)}^2 = \frac{1}{1-c_{13}}~, \quad c_{s(1)}^2=\frac{c_1-\frac{1}{2}c_{13}\bar{c}_{13}}{c_{14}(1-c_{13})}~,\nonumber\\
c_{s(0)}^2 = \frac{c_{123}(2 - c_{14})}{c_{14}(1 - c_{13})(2 + c_{13} + 3c_2)}~,
\end{align}
where \(c_{s(i)}\) is the  speed of the spin-\(i\) mode, \(c_{ij...}=c_{i}+c_{j}+...\) and \(\bar{c}_{13}=c_1-c_3\). Notice that when the only \(c_i\) that is nonzero is \(c_2\), the spin-1 mode has an indeterminate  speed and the spin-0 mode's speed diverges.

\subsection{Linear potential}
\label{linearpot}

Unlike the quadratic potential, the linear potential is qualitatively different than a general potential. In this case $\phi$ is not an auxiliary field in action \eqref{svtaction}, but a Lagrange multiplier,  and the equivalence with action \eqref{genaetheory} breaks down. However, the equivalence between action \eqref{svtaction} and the cuscuton action \eqref{cuscaction} still holds. A generic linear potential can be written as, 
	\begin{equation}
	V(\phi) = -\mu^2K_0\phi+M_{pl}^2\Lambda~,
	\end{equation}
where $K_0$ and $\Lambda$ are constants. Using this form  action \eqref{svtaction} can be written as,
\begin{align}
S[\phi,u,g]=&\int\mathrm{d}^4x\sqrt{-g}\Bigg[\frac{M_{pl}^2}{2}\left({}^{(4)}R-2\Lambda\right)\\
&
-\mu^2\phi\left(\nabla_\mu u^\mu-K_0\right)+\frac{M_{pl}^2}{2}\sigma\left(g_{\mu\nu}u^\mu u^\nu+1\right)\Bigg]~,\nonumber
\end{align}
after an obvious integration by parts. Variation with respect to $\phi$ yields the constraint $\theta=\nabla_\mu u^\mu=K_0$, while variation with respect to $u^\mu$ yields
\begin{equation}
\label{phisu}
-\mu^2 \nabla_\mu \phi=M_{pl}^2\sigma u_\mu\,.
\end{equation}
This equation implies that $u_\mu$ is hypersurface orthogonal when $\sigma\neq 0$,~\textit{i.e.} when the equivalence with the cuscuton is valid. Taking into account hypersurface orthogonality and the unit constraint, $\nabla_\mu u^\mu=K_0$ fully determines $u^\mu$, while Eq.~\eqref{phisu} can be seen as determining $\phi$. The stress-energy tensor for $u^\mu$ is
\begin{equation}
T_{\mu\nu}=M_{pl}^2\sigma (g_{\mu\nu} +u_\mu u_\nu)+\mu^2K_0\, \phi\, g_{\mu\nu}\,,
\end{equation}
the contracted Bianchi identity implies $\nabla^\mu T_{\mu\nu}=0$ and, after some manipulations, this yields
\begin{equation}
\label{sigmaeq}
h^{\mu\nu}\nabla_\mu\sigma+\sigma \, u^\mu\nabla_\mu u^\nu=0\,.
\end{equation}
 Eq.~\eqref{sigmaeq} can be seen as determining $\sigma$ and rendering the system closed. This condition is also a straightforward consequence of the fact that $u_{\mu}$ is hypersurface orthogonal by virtue of Eq.~\eqref{phisu}. 
Its geometric interpretation is the following: $u^\mu$ is the unit normal to a set of spacelike, constant mean curvature (CMC) hypersurfaces labelled by a time coordinate $T$. Eq.~\eqref{sigmaeq} implies that the acceleration of $u^\mu$ is also hypersurface orthogonal within each hypersurface of the $T$-foliation and hence $\sigma = $ constant surfaces provide a natural foliation of the $T = $ constant surfaces.

\section{Constraints and evolution equations}
\label{evolution}

We will now turn to determining the nature of the cuscuton's foliation. As discussed in the Introduction, in Ho\v{r}ava gravity the equations become second order in time derivatives only if one uses a specific foliation, and this singles out this foliation as being preferred. However, in the case of the cuscuton the equations are second
order even covariantly. Therefore any preferred foliation would 
have to be singled out more subtly. This is what we plan to clarify in this section.\

The existence of a preferred foliation can be demonstrated easily and clearly at perturbative level. It is instructive to inspect the kinetic term of the cuscuton in action \eqref{cuscaction} at quadratic order in field perturbations around a given background:
\begin{equation}
\sqrt{2X_{(2)}} = -\frac{1}{2}(2\,X_{(0)})^{-1/2} P^{00}_{(0)}\left(\partial_t\delta\phi\right)^2 + \dots
\end{equation}
Ellipsis denotes terms other than the kinetic term of $\delta\phi$, while \(P^{00}_{(0)}\) is the \(00\) component of $P^{\mu\nu}\equiv g^{\mu\nu}+(2X)^{-1} \nabla^\mu\phi\nabla^\nu\phi$ calculated for the background solution. 
 $X_{(n)}$ corresponds to $X$ expanded at $n$--th order in perturbations $\delta\phi$. In the cuscuton foliation the time coordinate is taken to be  $x^0=f(\phi)$. $P^{00}_{(0)}$ is zero in this foliation and positive in any other foliation. Hence,  the cuscuton perturbation is a ghost around every background in any  other foliation, and it becomes an elliptic mode in the cuscuton foliation, making the latter the preferred foliation. To clarify this point further a toy theory with two scalar fields, one of which exhibits the same behavior as the cuscuton, is discussed in detail in the Appendix. Note that the existence of an elliptic, instantaneous mode is expected in cuscuton theory, since the cuscuton perturbation is known to have infinite speed.
 
Here we have to make two clarifications. First of all this analysis has assumed that \(\mu^2>0\), as this is the standard cuscuton case and thus the focus of this paper.  For \(\mu^2<0\) the Hamiltonian remains unbound from below but this comes from the gradient terms which can not be removed by choice of foliation. The second clarification is that, from the perspective of effective field theory, one could choose to treat the cuscuton as a spurious degree of freedom coming from the truncation of a UV complete theory, provided that it's mass would be sufficiently high. The mass would then act as a cut-off and this would resolve the ghost problem. Though this is a possibility, it would effectively remove the cuscuton entirely as a degree of freedom within the range of validity of the effective field theory. This would go against the spirit of the original proposal.  To summarise, if one wishes to treat the cuscuton as a true degree of freedom then the theory only has a Hamiltonian which is bounded from below when studied in the cuscuton's foliation. \footnote{After our paper became publicly available on \url{arXiv.org} but prior to publication, Ref. \cite{Gomes:2017tzd} appeared, where a Hamiltonian analysis of the cuscuton is performed. We note that, provided that the cuscuton has a timelike gradient, one can always choose a foliation that renders it `homogeneous’ in the terminology of Ref. \cite{Gomes:2017tzd}. Taking this into account, the technical results of Ref. \cite{Gomes:2017tzd} agree with ours wherever there is overlap, even though there seem to be important differences in the physical interpretation.}

Next we turn our attention to the initial value formulation and consider a full, nonperturbative treatment. Constraint equations can be obtained by exploiting the generalized contracted Bianchi identity \cite{Barausse:2011pu,Jacobson:2011cc}.  Hence, we start with the derivation of this identity for this theory, by first considering the variation of the action generated by a diffeomorphism. Under infinitesimal transformations \(x^\mu \to x^\mu + \xi^\mu\), the fields transform by
\begin{align}
\delta g^{\mu\nu}&\equiv \mathcal{L}_{\xi}g^{\mu\nu}=-2\nabla^{(\mu}\xi^{\nu)},\nonumber\\
\delta u^{\mu}&\equiv\mathcal{L}_{\xi}u^{\mu}=\xi^{\nu}\nabla_{\nu}u^{\mu}-u^\nu\nabla_\nu\xi^\mu,\nonumber\\
\delta\phi&\equiv\mathcal{L}_{\xi}\phi=\xi^\nu\nabla_\nu\phi,
\end{align}
where \(\xi^\mu\) is a vector field which vanishes at infinity. As the action \eqref{svtaction} is diffeomorphism invariant, it will remain unchanged under the above variations,~\textit{i.e.}
\begin{align}
0=\int\mathrm{d}^4x\sqrt{-g}\left[-\left(M_{pl}^2E^\mu_{\;\;\nu}+u^\mu F_\nu\right)\nabla_\mu\xi^\nu\right.\nonumber \\
\left.\vphantom{-\left(2E^\mu_\nu+u^\mu_\nu\right)\nabla_\mu\xi^\nu}+\left\{\left(\nabla_\nu u^\mu\right)F_\mu+H\left(\nabla_\nu\phi\right)\right\}\xi^\nu\right],
\end{align}
where \(E_{\mu\nu}\), \(F_\mu\) and \(H\) were defined in Eqs.\eqref{equationsofmotion:1}-\eqref{equationsofmotion:3}, and all fields are taken to be off-shell at this stage. Integrating the first term by parts we are led to the contracted Bianchi identity,
\begin{equation}
\nabla_\mu\left(M_{pl}^2E^\mu_{\;\;\nu}+u^\mu F_\nu\right)+\left(\nabla_\nu u^\mu\right)F_\mu+H\nabla_\nu\phi=0,
\label{eq:contractedBianchi}
\end{equation}
where we observe that, recalling $\phi\neq$ constant, the diffeomorphism invariance renders the scalar equation of motion $H=0$ redundant. 
By choosing a set of coordinates $x^i$ with $i=1,2,3$ denoting the spacelike surfaces defined by constant $x^0$, we can expand the covariant derivatives in \eqref{eq:contractedBianchi} to give
\begin{equation}
\partial_0 \left(M_{pl}^2E^0_{\;\;\nu}+u^0 F_\nu\right) + \dots = 0\,,
\end{equation}
where the ellipsis denotes terms that do not involve more than two time derivatives of the fields. This expression implies that for the identity \eqref{eq:contractedBianchi} to hold, the combination $(M_{pl}^2E^0_{\;\;\mu}+u^0 F_\mu)$ cannot contain more than one time derivative and thus constitutes the four constraint equations of the theory. Moreover, once the fields on a given $x^0 = $ constant slice are considered on-shell, the contracted Bianchi identities ensure that
\begin{equation}
\partial_0 \left(M_{pl}^2E^0_{\;\;\nu}+u^0 F_\nu\right)=0\,,
\end{equation}
\textit{i.e.} the constraints are preserved by time evolution in all subsequent constant $x^0$ slices.
In covariant form, the four constraints can be written as
\begin{equation}
\hat{n}_\mu (M_{pl}^2E^\mu_{\;\;\nu}+u^\mu F_{\nu})=0\,,
\end{equation}
where $\hat{n}^\mu$ is the unit normal to the foliation, with \(\hat{n}_\mu \propto \delta^0_\mu\). 

Motivated by our perturbative considerations above, we will now choose the unit normal as $u_\mu$, {\em i.e.}~we will work in a foliation defined by the cuscuton. The vector equation of motion \eqref{equationsofmotion:2} gets trivially satisfied and the constraint equations can  be written in this foliation as,
\begin{equation}
u_\mu E^\mu_{\;\;\nu}=0\,.
\end{equation}
Further projecting this along the unit normal $u_\mu$ and onto the leaves with $h^\mu_{\;\;\nu}=\delta^\mu_\nu+u^\mu u_\nu$ yields the energy and momentum constraints, respectively. Recalling the form of the stress-energy tensor from Eq.\eqref{stressenergy}, and using the Gauss-Codazzi relations (see {\em e.g.}~\cite{Wald:1984rg}), one finds that the Hamiltonian constraint is
\begin{equation}\label{energy}
\frac{1}{2}\left(R-K_{ij}K^{ij}+K^2\right)-\frac{V(\phi)}{M_{pl}^2}=0,
\end{equation}
where $R$ is the scalar curvature corresponding to the induced $3$--metric on the leaves, while 
\begin{equation}
K_{ij}\equiv \frac{1}{2}\,\mathcal{\pounds}_u h_{ij}\,,
\label{extrinsiccurvature}
\end{equation}
is the extrinsic curvature. Similarly, the momentum constraint is obtained as
\begin{equation}\label{momentumfull}
\vec{\nabla}_iK^i_j-\vec{\nabla}_jK=0,
\end{equation}
where \(\vec{\nabla}\) is the covariant derivative on each slice associated with the induced metric $h_{i j}$. 
As we use the foliation defined by the scalar field, we have $\theta=K$, and the scalar equation of motion \eqref{equationsofmotion:3} implies that $K$ is a function of the scalar field, or the time coordinate in this setting. As a result, the momentum constraint can be simplified to
\begin{equation}\label{momentumsimple}
\vec{\nabla}_iK^i_j=0\,.
\end{equation}
As we have now fully specified the constraint equations, all that remains is to give the complete projection of the Einstein equations onto the leaves of the foliation. Once more recalling the form of the stress energy tensor \eqref{stressenergy} and using the standard Gauss-Codazzi relations one obtains,
	\begin{align}\label{dynamical}
	\mathcal{\pounds}_uK_{ij} = & - KK_{ij} + 2K_i^kK_{kj}+\frac{1}{M_{pl}^2}\left[V(\phi)-\frac{\mu^2\sqrt{2X}}{2}\right]h_{ij}\nonumber\\
	                            & + a_ia_j+\vec{\nabla}_ia_j-R_{ij},
	\end{align}
where $a_i\equiv h^\mu_i\,u^\nu \nabla_\nu u_\mu$ is the acceleration of $u_\mu$.

Picking a suitable gauge, Eq.~\eqref{dynamical} can then be interpreted as a first order time-evolution equation for the extrinsic curvature \(K_{ij}\) and the definition in Eq.~\eqref{extrinsiccurvature} then provides a first order time-evolution equation for the 3-metric. Since our foliation and gauge choice has rendered the scalar and the vector equation redundant or trivially satisfied, it is clear that the cuscuton has the same dynamical degree of freedoms as general relativity. However, things are slightly more subtle. $X$, which is present in the dynamical equations, is neither determined by them nor fixed by our gauge choices. I\textcolor{blue}{n}\textcolor{red}{f} fact, the trace of Eq.~\eqref{dynamical} is an elliptic equation for \(X\). Making the substitution \(\rho = 1/\sqrt{X}\), the trace of Eq.\eqref{dynamical} becomes
\begin{equation}
\vec{\nabla}^2\rho=\frac{\sqrt{2}}{\mu^2}\left[\frac{3\mu^4}{2\,M_{pl}^2}-\frac{\mathrm{d}^2V}{\mathrm{d}\phi^2}\right]-\frac{\rho V}{M_{pl}^2}+\rho K_{ij}K^{ij}.
\end{equation}
This equation is not a constraint and needs to be solved on each slice of the foliation. Hence, $X$ is an elliptic mode, which could be related to the lapse $N$ in this foliation and in a suitable gauge. 
The existence of the elliptic mode was already apparent in Section~\ref{sec:general-potential} where Eq.\eqref{hgenae} can be solved by using boundary conditions and it is in full agreement with the perturbative analysis we presented above.

Before closing this discussion, it is important to clarify the following: An elliptic equation that is not a constraint can be found in the initial value formulation of  vacuum general relativity when imposing a globally CMC foliation. In that case the trace of the dynamical equations becomes,
	\begin{equation}
	\label{ellgr}
\vec{\nabla}^2 N = N(K^2 + R) = N K_{i j}K^{i j}~,
	\end{equation}
where the second equality is obtained via the  Hamiltonian constraint. The interpretation in this case is that the existence of a CMC foliation is stronger than a gauge condition and, hence,  Eq.~\eqref{ellgr} arises as an existence condition for this foliation. This foliation is in no sense preferred and the elliptic mode is not physical, as in a different foliation the lapse function could be fully determined via a gauge choice.  The interpretation in the case of the cuscuton above is very different because the foliation we have chosen above is actually a preferred foliation, {\em i.e.}~the theory is not sensible in any other foliation, as has been shown  in the beginning of this Section. 

\section{Comments on spherical collapse}
So far, we have demonstrated that even though cuscuton theory is not equivalent to Ho\v{r}ava gravity, the theories do share some qualitative similarities; namely, there exists an instantaneous mode and a preferred foliation in both theories. In particular, this means that the definition of a black hole in the two theories is the same, and requires the existence of a universal horizon~\cite{Bhattacharyya:2015gwa}. One interesting question in such theories is whether these horizons actually form dynamically. In the case of Ho\v{r}ava gravity there is some indication that they can; see~\emph{e.g.},~\cite{Garfinkle:2007bk} as well as~\cite{Bhattacharyya:2015uxt} for a recent re-interpretation of some of the results. Ref.~\cite{Saravani:2013kva} has instead studied the dynamical formation of universal horizons in cuscuton theory, under certain simplifying approximations.

In particular, Ref.~\cite{Saravani:2013kva} worked in the decoupling limit of the theory, where the cuscuton's back-reaction on the spacetime geometry is considered to be negligible. One then has to solve the cuscuton's equation of motion on a spacetime that solves Einstein's equations. The authors of Ref.~\cite{Saravani:2013kva} focused on a simplified collapse scenario, that of a freely falling,  spherical, thin shell of dust. In this approximation, as a consequence of Birkhoff's theorem, the spacetime exterior and interior of the shell are Schwarzschild and Minkowski spacetimes respectively. Solving equation~\eqref{equationsofmotion:3} (equivalently~\eqref{scalarequation}) for the cuscuton in the decoupling limit then amounts to finding the appropriate CMC foliation of this background, \emph{i.e.} a CMC foliation that  covers both the interior and the exterior of the shell, and is suitably matched at the shell radius. In Ref.~\cite{Saravani:2013kva} this route was taken by constructing the corresponding unit timelike vector respecting spherical symmetry (playing the role of the unit normal to the cuscuton foliation). 

As a further approximation, the cuscuton potential was assumed to be constant. Within this approximation a universal horizon is shown to eventually form. This result is certainly an indication that universal horizons can form from gravitational collapse in cuscuton theory.  Some degree of caution is warranted though, due to the large number of approximations employed. 

For instance, the decoupling limit allows for asymptotically flat solutions which do not actually exist in full cuscuton theory, as discussed previously. As also argued in Ref.~\cite{Saravani:2013kva}, one does not expect the asymptotic behavior of the metric to be important at small scales, where the universal horizon  eventually appears because of separation of scales. However, this statement does need to be taken with a pinch of salt, both because of the global nature of the causal structure and the fact that the theory has infinitely fast propagation. 

The use of a constant potential is also quite restrictive. As we have shown in Section \ref{formaleq}, the theory with  a linear or constant potential differs significantly --- at least generically --- from the theory with a more general potential. Moreover, the assumption of constant $V(\phi)$ implies that $\phi$ appears in Eq.~\eqref{scalarequation} in the combination $\nabla^\mu\phi/\sqrt{2X}$ only. When one further employs the decoupling limit, $\phi$ drops out of Einstein's equations and the theory acquires an accidental invariance under $\phi \to \tilde{\phi}(\phi)$. This symmetry is clearly not present in general. Since $\phi$ plays the role of time in the preferred foliation, absence of this symmetry means that the foliation defined by $\phi$ is uniquely labelled (for a general potential $\phi$ is not even shift-symmetric) and, hence, needs to satisfy stricter regularity requirements. 

Before closing this section we would like to clarify one more subtle point. The solution found in Ref.~\cite{Saravani:2013kva} for \(V(\phi)= \mathrm{constant}\) seems to be converging asymptotically in time to the static solutions of Einstein-\aether theory with only \(c_2\) nonzero \cite{Berglund:2012bu,Bhattacharyya:2014kta}, for which the universal horizon is located at \(r=1.5M\). As discussed in Section \ref{linearpot}, the cuscuton with linear  or constant potential is not equivalent to some corner of Einstein-\aether theory. The explanation for this apparent coincidence is the following. With only \(c_2\neq 0\) the stress-energy tensor of Einstein-\aether theory is
\begin{equation}
T_{\mu\nu}=\left(c_2\nabla_\alpha\left[Ku^\alpha\right]-c_2\frac{K^2}{2}\right)g_{\mu\nu}+\sigma u_\mu u_\nu.
\end{equation}
 As we have argued in Section \ref{sec:general-potential}, static solutions have $\sigma=0$. The solutions found in Refs.~\cite{Berglund:2012bu,Bhattacharyya:2014kta} are static and have $K=0$, so they have vanishing stress-energy tensor. Additionally, they are spherically symmetric and hence the aether is hypersurface orthogonal. 
In summary, the metric is a static, spherically symmetric solution to vacuum Einstein equations and the aether in a hypersurface orthogonal vector that has zero divergence. These are precisely the same assumptions and equations that are used in Ref.~\cite{Saravani:2013kva}.

\section{Discussion}
Ho\v{r}ava gravity is a  prototypical example of a theory with a preferred foliation and an elliptic, instantaneous mode that persists even at low energies. What makes a certain foliation preferred in this theory is that the field equations have higher order than second time derivatives in every other foliation.  In this work, cuscuton theory was used as an explicit example of a theory where the field equations are second order in~\emph{every} foliation and yet there is both an elliptic mode and a preferred foliation. Therefore, one of the takeaway messages of this work is that 
 preferred foliation may arise in  diverse and even subtle manners in different theories.

Towards establishing the above, we started out by reviewing a known reformulation of cuscuton theory as a scalar-vector-tensor theory. In this reformulation either the vector or the scalar can be seen as an auxiliary field. Depending on whether one solves for the scalar or for the vector, and under certain mild assumptions, one ends up with either a special case of (generalized) Einstein-{\ae}ther theory or cuscuton theory. 
This demonstrates dynamical equivalence between the latter two. Interestingly,  in this special corner of Einstein-{\ae}ther theory the {\ae}ther is hypersurface orthogonal simply by virtue of the equations of motion. Needless to say, this behavior is not generic for Einstein-{\ae}ther theory. 
The equivalence between cuscuton theory and this special case of Einstein-{\ae}ther theory is somewhat delicate. In particular, it is quite sensitive to the form of the cuscuton potential and theories with generic potentials can be mapped to a generalized version of Einstein-{\ae}ther theory. Additonally, the equivalence appears to break down if stationarity is assumed.

Our analysis has clearly shown that  cuscuton theory is not equivalent to (a special corner of) Ho\v{r}ava gravity, as has been claimed in the literature. The fact that the cuscuton always has second order equations of motion is perhaps sufficient to convince oneself that such a claim could not have been true; nevertheless our demonstration hopefully settles the issue explicitly. 

Nonetheless, this is not to say that cuscuton theory does not bear qualitative similarities to Ho\v rava gravity. We have performed a preliminary analysis of the initial value formulation in cuscuton theory.  This allowed us to group the field equations as constraints and evolution equations. In the process, we have shown that in the cuscuton's foliation there exists also an elliptic equation that is not a constraint, much like Ho\v rava gravity. This is a nonperturbative manifestation of an instantaneous mode and in agreement with the fact that the cuscuton appears to have infinite speed at perturbative level. It was further shown that the cuscuton's foliation is singled out as preferred by the fact that a cuscuton perturbation becomes a ghost in any other foliation.

Our results show that the causal structure of cuscuton theory is quite similar to that of Ho\v{r}ava gravity but there are also crucial differences in their dynamics. In the context of black hole physics, this implies that both theories are expected to have universal horizons but the dynamical formation of such horizons does not have to take place in the same fashion.

\begin{acknowledgments}
The work of AEG is supported by the Science and Technologies Facilities Council Grant No: ST/L00044X/1.
The research leading to these results has received funding from the European Research Council under the European Union's Seventh Framework Programme (FP7/2007-2013) / ERC Grant Agreement n. 306425 ``Challenging General Relativity''. AEG would like to thank University of Nottingham for their warm hospitality during the course of this work.
\end{acknowledgments}
\appendix

\label{app}

\section{A simpler second order theory with preferred foliation}
The goal of this appendix is to present an example of a toy theory of a pair of scalar fields, where the equations of motion are  second order in any foliation, but there  still exists a preferred foliation. 

Consider the  theory
	\begin{equation}\label{lag:toy-model}
	\mathcal{L} = \frac{1}{2}(u^\mu\nabla_\mu\phi)^2 - \frac{1}{2}h^{\mu \nu}(\nabla_\mu\psi)(\nabla_\nu\psi) - \lambda\phi\psi~,
	\end{equation}
where $\phi$ and $\psi$ are a pair of scalar fields, $u^\mu$ is unit timelike and hypersurface orthogonal, $h^{\mu \nu} = g^{\mu \nu} + u^\mu u^\nu$, and $\lambda$ is a coupling between the fields. 
For the purposes of this discussion, we disregard the dynamics of $u^\mu$ and the metric and we treat them as background field. For simplicity we will further assume that the metric is flat and that $u^\mu$ is constant. However, rendering  $u_\mu$ and the metric dynamical would not affect our conclusion provided that $u^\mu$ is restricted to be unit timelike and hypersurface orthogonal.

Upon extremizing the action~\eqref{lag:toy-model}, the equations of motion for the fields are
	\begin{equation}\label{EOM:phiSF}
	\nabla_\mu[u^\mu u^\nu\nabla_\nu\phi] = -\lambda\psi~, \quad \nabla_\mu[h^{\mu \nu}\nabla_\nu\psi] = \lambda\phi~.
	\end{equation}
Adapting to the foliation defined by the vector field such that $u_\mu = -\nabla_\mu t$ where $t$ is some choice of Minkowski time, the equations of motion~\eqref{EOM:phiSF} simplify to
	\begin{equation}\label{EOM:phiSF:M}
	\ddot{\phi} = -\lambda\psi~, \quad \vec{\nabla}^2\psi = \lambda\phi~,
	\end{equation}
where an overdot denotes a partial derivative with respect to $t$ and $\vec{\nabla}^2$ is the standard Laplacian on each such leaf.

For $\lambda = 0$, the equations of motion for $\phi$ and $\psi$~\eqref{EOM:phiSF:M} decouple Integrating the $\phi$ equation, one has $\phi(t, \boldsymbol{x}) = \varpi(\boldsymbol{x})t + \phi^0(\boldsymbol{x})$, where $\boldsymbol{x}$ denote a set of spatial coordinates on each $t = $ constant leaf, and the functions $\phi^0(\boldsymbol{x})$ and $\varpi(\boldsymbol{x})$ are the initial values of the field $\phi$ and its first time derivative respectively. On the other hand, $\psi$ becomes manifestly elliptic and it's profile is determined by solving a Laplace equation subject to suitable boundary conditions. 

For $\lambda \neq 0$, the time evolution can be described qualitatively as follows: upon introducing a function $\pi = \dot{\phi}$, the equations~\eqref{EOM:phiSF:M} can be cast in a first order form as follows
	\begin{equation}\label{EOM:phiSF:M:1stO}
	\dot{\pi} = -\lambda\psi~, \quad \dot{\phi} = \pi, \quad \vec{\nabla}^2\psi = \lambda\phi~.
	\end{equation}
Given initial data consisting of the profiles of $\pi$, $\phi$ and $\psi$ on some initial slice at $t = t_0$, one may in principle integrate the first two equations of~\eqref{EOM:phiSF:M:1stO} in time,~\emph{e.g.} over a time step $\epsilon$, to construct the $\pi$ and $\phi$ profiles on the $t = t_0 + \epsilon$ slice. However, to carry the process out for subsequent times, one also needs to solve the elliptic equation for $\psi$ on the $t = t_0 + \epsilon$ slice, whose profile would then feed the evolution of $\pi$ and so on. In other words, the elliptic equation for $\psi$ needs to be solved on~\emph{every} time slice and it ultimately describes the proper evolution of a physical degree of freedom. Hence,   the system contains an elliptic equation that is not preserved by time evolution. One may contrast the situation here with the more familiar context of general relativity where elliptic constraint equations do arise but are only needed to be solved on the initial time slice.

In the discussion above we have adopted a specific foliation, so one might have a reasonable concern that our conclusions might be affected by this choice. We will now show that this is the only sensible foliation, and hence it is a preferred foliation. 
To this end, consider  another inertial frame which is everywhere constantly boosted with respect to $u^\mu$. If $\eta$ denotes the boost parameter, then $u^\mu = \{\cosh\eta,~\sinh\eta\,\Omega^i\}_{i = 1}^3$ in the new frame, where $\Omega^i$ are constants constrained by $\sum_{i = 1}^3(\Omega^i)^2 = 1$. If we use an overdot to denote a time derivative in this new frame and a subscript $i$ to denote the spatial derivative with respect to $x^i$, then the Lagrangian~\eqref{lag:toy-model} in this new frame reads
	\begin{equation*}
	\begin{split}
	\mathcal{L} = & \frac{1}{2}\cosh^2\eta\,\dot{\phi}^2 - \frac{1}{2}\sinh^2\eta\,\dot{\psi}^2 + \frac{1}{2}\sinh^2\eta\,(\Omega^i\phi_i)^2 \\
	              & + \sinh\eta\cosh\eta\,(\dot{\phi}\phi_i - \dot{\psi}\psi_i)\Omega^i \\
	              & - \frac{1}{2}(\delta^{i j} + \sinh^2\eta\,\Omega^i\Omega^j)\psi_i\psi_j - \lambda\phi\psi~.
	\end{split}
	\end{equation*}
The momenta conjugate to $\phi$ and $\psi$ are then
	\begin{equation}\label{momenta:FS}
	\begin{split}
	\pi_{\phi} & = \cosh\eta\,(\cosh\eta\,\dot{\phi} + \sinh\eta\,\Omega^i\phi_i)~, \\
	\pi_{\psi} & = -\sinh\eta\,(\sinh\eta\,\dot{\psi} + \cosh\eta\,\Omega^i\psi_i)~,
	\end{split}
	\end{equation}
leading to the following expression for the Hamiltonian
	\begin{equation*}
	\begin{split}
	\mathcal{H} = & \frac{1}{2}\cosh^2\eta\,\dot{\phi}^2 - \frac{1}{2}\sinh^2\eta\,\dot{\psi}^2 - \frac{1}{2}\sinh^2\eta\,(\Omega^i\phi_i)^2 \\
	              &  + \frac{1}{2}(\delta^{i j} + \sinh^2\eta\,\Omega^i\Omega^j)\psi_i\psi_j + \lambda\phi\psi~,
	\end{split}
	\end{equation*}
where $\dot{\phi}$ and $\dot{\psi}$ are implicitly given in terms of $(\pi_{\phi}, \phi_i)$ and $(\pi_{\psi}, \psi_i)$ respectively, according to~\eqref{momenta:FS}. The Hamiltonian in the above form is written in terms of sums and differences of perfect squares (modulo the interaction term). $\mathcal{H}$ is bound from below only for  $\eta = 0$. Hence,  the choice $\eta = 0$ is the only sensible one and this single out the the foliation defined by $u_{\mu}$, in which $\psi$ behaves like an elliptic mode (note that the $\psi$ momenta vanishes for $\eta = 0$). In this foliation time evolution can be consistently formulated (albeit in a nonrelativistic manner), whereas in any other foliation it straightforwardly follows that $\psi$ would be ghost. It should also be noted that, choosing the opposite sign for the \(h^{\mu\nu}\nabla_\mu\psi\nabla_\nu\psi\) term  would be analogous to  the \(\mu^2<0\) case in the beginning of section \ref{evolution}. There would still be an instability, but it would be a gradient instability rather than a ghost-like instability.


\end{document}